\documentclass[11pt,dvips]{article}
\usepackage{graphicx}
\usepackage{amsmath}  
\usepackage{bm} 

\def\ds{\displaystyle}
\def\ni{{\noindent}}
\def\di{{\partial}}
\def\ie{{\it i.e.~}}

\def\be{\begin{equation}}
\def\ee{\end{equation}}
\def\bea{\begin{eqnarray}}
\def\eea{\end{eqnarray}}

\def\L1{{\cL_{(1)}}}

\def\cL{{\cal L}}

\def\nn{{\nonumber}}

\usepackage{amsmath}
\usepackage{amssymb}
\usepackage{indentfirst}
\usepackage{verbatim} 
\usepackage{graphicx}
\usepackage{slashed}
\usepackage{verbatim}
\usepackage{fancyhdr}
\usepackage{rotating}
\def\bra{{\langle}}
\def\ket{{\rangle}}
\def\e{\textrm{e}}
\def\F!{\;\;\;\;\;\;\,\,}
\def\H!{\;\;\;\,}
\def\cL{{\cal L}}      
\def\cH{{\cal H}}      

\def\ds{\displaystyle}
\def\di{{\partial}}
\def\xp{{{ x}^{\prime}}}
\def\xpp{{{ x}^{\prime\prime}}}
\def\xppp{{{ x}^{\prime\prime\prime}}}
\def\zp{{{ z}^{\prime}}}
\def\zpp{{{ z}^{\prime\prime}}}

\def\x{{\bf x}}        
 
\def\r{{\bf r}}

\def\p{{\bf p}}
\def\q{{\bf q}}

\def\nn{\nonumber}
\def\cY{{\cal Y}}           
\def\cC{{\cal C}}           
\def\cQ{{\cal Q}}
\def\cF{{\cal F}}
\def\cG{{\cal G}}
\def\de{\delta}
\def\tx{\textrm}
\def\E1{\textrm{E}_{1}}     

\def\ub{{\bar u}}
\def\vb{{\bar v}}
\def\psib{{\bar \psi}}

\begin{document}
\title{Effect of Virtual Pairs on the Inter-quark Potential} %
\author{A. Chigodaev and J. W. Darewych  \cr  
{\it Department of Physics and Astronomy,}\cr
{\it York University, Toronto, ON M3J 1P3, Canada}}
\date{\small September 9, 2013}
\maketitle
\begin{abstract}
We use the variational method, in a reformulated Hamiltonian formalism of QCD, to derive the wave equation 
for a heavy quark-antiquark system using a trial state that contains a component with a virtual light 
quark pair. We examine the quark-antiquark potential in the non-relativistic limit using an approximate trial 
ground-state wave function. We find that the potential exhibits a confining character due to the 
inclusion of the virtual pair component in the trial state. 
\end{abstract}

\baselineskip= 18pt 
\def\Aslash{A\kern-0.47em/}
\def\aslash{a\kern-0.47em/}
\def\overc{\overline c}
\def\overb{\overline b}
\def\overt{\overline t}
\def\overq{\overline q}
\def\ds{\displaystyle}

\def\di{\partial}
%
%
%
\renewcommand{\theequation}{1-\arabic{equation}}  
\setcounter{equation}{0}
\section{Introduction}
The problem of deriving an analytic (or semi-analytic) expression for the 
 the force between quarks in an ab-initio way continues to be the subject of investigation 
(see, for example, the book by J. Greensite  \cite{Greensite} and the 
Quark Confinement and Hadron Spectrum conferences and corresponding proceedings 
http://www.confx.de/index.html).
The form of the potential between a fixed quark and antiquark, separated by a distance $r$
has been calculated in lattice gauge theory \cite{Greensite, Knechtli}. 
The shape of the potential is found to resemble the phenomenological so-called Cornell potential $ \ds V(r) = - \frac{\alpha}{r} + b \,
r$ \cite{Cornell 1, Cornell 2}.

\par
Our aim in this work is to determine the form of the potential in heavy quarkonium, by using the variational method, in a reformulated Hamiltonian formalism of QCD. Such an approach has been found to be useful for the study of few-body bound states
 in QFT, including QED \cite{TerDar, BarhamDar} as well as model scalar theories with non-linear mediating fields 
 \cite{DarDuv2, ChigoDar1}.
%
%
\section{QCD Lagrangian and Reformulation}
The QCD Lagrangian density (units: $\hbar = c = 1$), suppressing the spinor and flavour indices, is ~\cite{RPP, PascTarr, DGH}
\begin{align}
  \cL = \, - \frac{1}{4} \, (F_{\mu\nu}^a)^2 - \frac{1}{2 \, \xi} (\di^\mu A_\mu)^2 + \psib_i(i \slashed{D}_{ij} - m) \, \psi_j, 
  \label{EQ:LQCD1}
\end{align}
where the shorthand definitions are 
\begin{gather}
  F_{\mu\nu}^a = \di_\mu \, A_\nu^a - \di_\nu \, A_\mu^a + g_s \, f^{abc} \, A_\mu^b \, A_\nu^c, \\
  (D_\mu)_{ij}  = \di_\mu \de_{ij} - i \, g_s A_\mu^a \, T^a_{ij}. 
\end{gather}
The quark field $\psi_i$ is a Dirac spinor where the index $i = 1,2,3$ is a colour index. The vector boson field $A^a_\mu$ represents gluons carrying a Lorentz index $\mu = 0,..., 3$ and a colour index of the adjoint representation $a = 1,...,8$.  The dimensionless coupling constant $g_s$ characterizes the strength of the strong interaction. $F_{\mu\nu}^a$ is the non-Abelian field strength tensor, while $D_\mu$ is the covariant derivative. 
The eight $SU(3)$ group generators $T^a_{ij}$ and the (completely antisymmetric)  structure constants $f^{abc}$ obey the usual commutation relation.
\par
The term containing $\xi$ is known as the gauge fixing term. Physical observables do not, in principle, depend on the value of $\xi$. However, the canonical quantization of the gauge field is problematic in its absence since the conjugate momentum is undefined, i.e. $\pi^0 = \ds\frac{\partial \cL}{\partial \dot{A}_0} = 0$ if $\xi = \infty$. 

\par
It is convenient to express the Lagrangian density (\ref{EQ:LQCD1}) in the following form
\begin{align}
  \cL = \cL_A + \cL_\psi + \cL_{\psi A} + \cL_{3A} + \cL_{4A},
  \label{EQ:LQCD2}
\end{align}
where 
\begin{align}
  \cL_A &= - \frac{1}{4} (\di_\mu A_\nu^a - \di_\nu A_\mu^a) (\di^\mu A^{\nu a} - \di^\nu A^{\mu a}) - \frac{1}{2 \, \xi} (\di^\mu \, A^a_\mu)^2 \label{EQ:LA}, \\
& \simeq \; \frac{1}{2}A_{\mu}^a\left(\di^2 g^{\mu\nu} - \di^{\mu} \di^{\nu}\right) A_{\nu}^a + \frac{1}{2\xi} A_{\mu}^a\di^{\mu}\di^{\nu}A_{\nu}^a, \label{FFL2}
\end{align}
where  $\di^2 = \ds\frac{\di^2}{\di t^2} - \nabla^2$,  
$g^{\mu\nu} = \textrm{diag}[1, -1, -1, -1]$, $\simeq$ means
equivalence, modulo an irrelevant total derivative (surface) term, and
\begin{gather}  
\cL_\psi =  \psib (i\, \slashed{\di} - m) \psi,  \label{EQ:LPSI} \\
  \cL_{\psi A} = g \, \psib_i \, \slashed{A}^a T^a_{ij} \psi_j, \label{EQ:LPA}\\
  \cL_{3A} = - g_s\, f^{abc} (\di_\mu A^a_\nu) \, A^{\mu \, b} \, A^{\nu \, c}, \label{EQ:L3A} \\
  \cL_{4A} = - \frac{1}{4} \, g_s^2 (f^{abc} A_\mu^b A_\nu^c) (f^{ade} A^{\mu d} A^{\nu e}) \label{EQ:L4A}.
\end{gather}
\par
The equation of motion for the gauge field $A_\mu^a$ that follows from (\ref{EQ:LQCD2}) is
\begin{equation}
  \left(\di^2 g^{\mu\nu} - (1 - \frac{1}{\xi}) \, \di^{\mu}\di^{\nu}\right) A_{\nu}^a = \rho^{\mu \, a} (x), 
  \label{EQ:EM_A}
\end{equation}
where the ``source'' $\rho^{\mu \, a} (x)$  of this inhomogeneous equation is 
\begin{align}
  \rho^{\mu \, a}(x)  = & - g_s \, \psib_i(x) \, \gamma^\mu \,  T^a_{ij} \, \psi_j(x) + g_s \, f^{abc} \di^\nu \, \Big(A^b_\nu(x) \,  A^{\mu c}(x) \Big) \nn \\
  & - g_s \, f^{abc} \, \Big(\di^\mu A^{\nu b}(x) - \di^\nu A^{\mu b}(x)\Big) \,  A^c_\nu(x) + g_s^2 \, f^{abc} \, f^{cde} A^b_\nu(x) \,  A^{\mu d}(x) \, A^{\nu e}(x).  \label{EQ:source}
\end{align}
Note that in the QED case (\ie U(1), cf. \cite{TerDar}) the cubic (\ref{EQ:L3A}) and quartic (\ref{EQ:L4A}) terms do not arise, and the source term would contain the fermion fields only, that is, just the first term  of equation (\ref{EQ:source}). 
\par
Equation (\ref{EQ:EM_A}) can be written in integral from: 
\begin{align}
  A^a_\nu(x) = \int d\xp \, \Delta^{ab}_{\mu\nu}(x - \xp) \, \rho^{\nu b}(\xp), 
  \label{EQ:FORMAL_SOL}
\end{align}
where $\Delta^{ab}_{\mu\nu}(x - \xp)$ is the Green function of the homogeneous equation, that is 
\begin{align}
  \left(\di^2 g^{\mu\nu} - (1 - \frac{1}{\xi}) \, \di^{\mu}\di^{\nu}\right) \Delta_{\nu\rho}^{ab}(x-y) = \de^{ab}\de(x-y) \de^\mu_{\, \rho}.
\end{align}
In practice, one must specify a gauge. We shall choose the Feynman gauge, where $\xi = 1$, for which the Green function takes on the simple form: 
\begin{align}
  \Delta_{\mu\nu}^{ab}(x-y) = - \int\frac{d^4k}{(2\pi)^4} \, \frac{\de^{ab} \, g_{\mu\nu} \, \e^{-i k\cdot(x-y)}}{k^2 + i \epsilon}.
  \label{EQ:G_PROP}
\end{align}
\par
Substitution of the ``formal solution" (\ref{EQ:FORMAL_SOL}) into the Lagrangian density (\ref{EQ:LQCD2}), 
yields, after some algebra, the reformulated Lagrangian density
\begin{align}
  \cL = \cL_\psi + \cL_{\psi A}^R + \cL_{3A}^R + \cL_{4A}^R, 
  \label{EQ:LQCD3}
\end{align}
where, the reformulated terms (denoted by the superscript $R$) are
\begin{align}
  \cL_A + \cL_{\psi A} & \rightarrow \cL_{\psi A}^{R} = - \frac{1}{2} \, g_s^2 \, \psib \, \gamma^\mu T^a \psi \int d\xp \Delta^{ab}_{\mu\nu} (x - \xp) \, \rho^{\nu b}(\xp), \\
  \cL_{3A} & \rightarrow \cL_{3A}^R =  \, - \, g_s \, f^{abc}  \int d\xp \, \di_\mu \, \Delta^{ad}_{\nu\sigma}(x - \xp) \, \rho^{\sigma d}(\xp) \nn \\
  & \F!\F! \times \int d\xpp \Delta^{\mu\alpha \, be}(x - \xpp) \, \rho^{e}_{\alpha}(\xpp) \int d\xppp \Delta^{\nu\beta \, cf}(x - \xppp) \, \rho^{f}_{\beta}(\xppp), \\
  \cL_{4A} & \rightarrow \cL_{4A}^R = - \frac{1}{4} \, g_s^2 \left(f^{abc} \int d\xp \Delta^{bi}_{\mu\sigma}(x - \xp) \, \rho^{\sigma i}(\xp) \int d\xpp \Delta^{cj}_{\nu\tau}(x - \xpp) \, \rho^{\tau j}(\xpp)\right) \nn \\ 
  & \F!\F! \times \left(f^{ade} \int d\zp \Delta^{\mu\alpha \, dk}(x - \zp) \, \rho^{k}_{\alpha}(\zp) \int d\zpp \Delta^{\nu\beta \, el}(x - \zpp) \, \rho^{l}_{\beta}(\zpp)\right). 
\end{align}
\par
The purpose of reformulation is to eliminate the mediating field from the Lagrangian density while preserving its effects through its propagator. For theories with linear mediating fields, such as QED, this is easily achieved, since the ``source" 
$\rho$ does not contain the mediating photon field. However, for QCD the ``source" term (\ref{EQ:source}) does contains 
the gauge fields $A_\mu^a$ so it is not possible to solve (\ref{EQ:EM_A}) to express the gluon fields in terms of the quark fields explicitly. Hence we must resort to approximations; an obvious approach is iteration. 
\par
In lowest iterative order, which is what we shall use, the ``source'' term is truncated to
\begin{align}
  \rho_\mu^a(x) = - g_s \, \psib_i(x) \gamma_\mu T^a_{ij} \psi_j(x).
  \label{EQ:SOURCE_T}
\end{align}
 Using this lowest-order truncation in the terms of the Lagrangian density (\ref{EQ:LQCD3}) leads to
\begin{align}
  \cL_{\psi A}^R = & \, + \frac{1}{2} \, g_s^2 \, \psib(x) \, \gamma^\mu T^a \psi(x) \int \frac{d\xp \, dk}{(2\pi)^4} \, \frac{\e^{ -i k\cdot(x - \xp)}}{k^2} \, \psib(\xp) \, \gamma_\mu T^a \psi(\xp), 
  \label{EQ:INTQCD1} \\
  \cL_{3A}^R = & \, - i \, g_s^4 \, f^{abc} \int d\xp \, d\xpp \, d \xppp \, \frac{dk \, dq \, dp}{(2\pi)^{12}} \, \frac{\e^{- i k\cdot(x - \xp)}}{k^2} \, \frac{\e^{- i q\cdot(x - \xpp)}}{q^2} \, \frac{\e^{- i p\cdot(x - \xppp)}}{p^2} \nn \\
  & \F!\H! \times \bigg\{\psib(\xp) \, \gamma_\nu T^a \psi(\xp) \bigg\} \,\bigg\{k_\mu \, \psib(\xpp) \, \gamma^\mu T^b \psi(\xpp) \bigg\}\, \bigg\{\psib(\xppp) \, \gamma^\nu T^c \psi(\xppp)\bigg\}, \\
  \cL_{4A}^R =  & - \frac{1}{4} \, g_s^6 \, f^{abc} f^{ade} \int d\xp \, d\xpp \, d\zp \, d\zpp \, \frac{dk \, dq \, dp \, dl}{(2\pi)^{16}} \; \frac{\e^{- i k\cdot(x - \xp)}}{k^2} \, \frac{\e^{- i q\cdot(x - \xpp)}}{q^2} \nn \\
  & \F!\H! \times\frac{\e^{- i p\cdot(x - \zp)}}{p^2} \, \frac{\e^{- i l\cdot(x - \zpp)}}{l^2} \; \bigg\{ \psib(\xp) \, \gamma_\mu T^b \psi(\xp)\bigg\} \, \bigg\{\psib(\xpp) \, \gamma_\nu T^c \psi(\xpp) \bigg\} \nn \\
  & \F!\H! \times\bigg\{\psib(\zp) \, \gamma^\mu T^d \psi(\zp) \bigg\} \, \bigg\{\psib(\zpp) \, \gamma^\nu T^e \psi(\zpp) \bigg\}.
  \label{EQ:REF_LQCD}
\end{align}
This reformulated Lagrangian density contains only quark fields; the interactions involving the mediating gluon field are represented by the gluon propagators. 
Note that $\cL^R$ contains terms corresponding to one gluon ($\cL_{\psi A}^R$), three gluon ($\cL_{3A}^R$) and four gluon 
($ \cL_{4A}^R$) interactions. 
Summation of the colour indices is implied, {\it i.e.} $\psib \gamma^\mu T^a \psi \equiv \psib_i \gamma^\mu T^a_{ij} \psi_j$.
\par
The Hamiltonian density corresponding to the Lagrangian density (\ref{EQ:LQCD3}) follows from the usual expression 
\begin{align}
  \cH_R = \Pi_{\psi} \dot{\psi} + \dot{\psib} \, \Pi_{\psib} - \cL_R,
  \label{EQ:HAM}
\end{align}
where, the conjugate momenta are defined in the usual way:
\begin{align}
  \Pi_{\psi} = \frac{\di \cL}{\di \dot{\psi}} = i \, \psib \, \gamma^0, \F!   \Pi_{\psib} = \frac{\di \cL}{\di \dot{\psib}} = - i \, \gamma^0 \, \psi.
\end{align}
%
%
\par

\section{Quantization}
In the quantized theory the fields are operators that satisfy anti-commutation rules that are very similar to those of QED, but with the inclusion of colour indices (and, flavour indices if needed). To specify our notation, we quote the anti-commutation relations for given flavour:
\begin{align}
  \big\{a_{s,i}(\p), a^{\dagger}_{\sigma,j}(\q)\big\} =   \big\{b_{s,i}(\p), b^{\dagger}_{\sigma,j}(\q)\big\} = \de(\p - \q) \, \de_{s\sigma} \, \de_{ij}.
\end{align}
Here, $a_{s,i}(\p)$ and $a^{\dagger}_{s,i}(\p)$ are quark annihilation and creation operators, respectively, with spin and 
colour indices $s$ and $i$. Similarly, $b_{s,i}(\p)$ and $b^{\dagger}_{s,i}(\p)$ are the corresponding anti-quark operators. 

In terms of these operators, after integrating out the spatial coordinates and normal-ordering the ladder operators to remove the infinite vacuum energy, one obtains the Hamiltonian operator 
\begin{gather}
  H_R = \int d\x \, : \cH(x) : \,  =  H_\psi - \int d\x \, : \Big(\cL_{\psi \, A}^R + \cL_{3A}^R + \cL_{4A}^R \Big) : \;  \equiv H_\psi + H_{\psi \, A}^R + H_{3A}^R + H_{4A}^R
  \label{EQ:QCDHAMREFORM}
\end{gather}
where the free Hamiltonian is given by
\begin{align}
  H_\psi = \sum_{i}^{3} \sum_{s = \pm} \int d\p \; \omega_\p \, \bigg\{ a^{\dagger}_{s, i}(\p) \, a_{s,i}(\p) + b^{\dagger}_{s,i}(\p) \, b_{s,i}(\p) \bigg\},  \label{EQ:QCDHAMFREE}
\end{align}
and $\omega_\p = \sqrt{\p^2 + m^2}$, with $m$ being the mass of the quark (of given flavour).
Evidently, this is like the expression for QED, except for the extra sum over the colour index.
\par
Unfortunately, expressions for the interactions terms $H_{\psi \, A}^R$, $H_{3A}^R$ and $H_{4A}^R$, when written in terms of the quark creation and annihilation operators 
are not at all simple and will not be written out explicitly. 
Instead, their matrix elements, in the context of a definite trial state, will be provided. 
\par
For the study of stationary bound states of quarks, 
we will switch from the present interaction picture to the Schr\"odinger picture by means of the unitary transformation 
$|\Psi_I\ket = e^{i H_0 t} |\Psi_S\ket$, where $H_0$ is the 
free Hamiltonian $H_\psi$ of equation (\ref{EQ:QCDHAMREFORM}). 
Henceforth, all matrix elements will be written in the Schr\"odinger picture.
%
%
\section{Multi-Component Quark-Antiquark (Meson) State in QCD}
It is not possible to determine exact eigensolutions of the QCD eigenvalue equation $H_R |\Psi\ket = E |\Psi\ket$. 
We shall use the variational method to study approximate solutions. For a quark-antiquark system we shall use the 
following trial state 
\begin{align}
  | \Psi_{t} \ket = C_F \, | \Psi_{2} \ket + C_G \, | \Psi_{4} \ket,
  \label{EQ:QCDTRIAL24}
\end{align}
where, the two Fock-space components are
\begin{align}
  | \Psi_2 \ket = & \sum_{i,j} \, \Omega_{ij} \sum_{\kappa, \lambda} \int d\p_{1,2} \, \cF_{\kappa, \lambda}(\p_{1,2}) \, a^{\dagger}_{\kappa,i, a}(\p_1) \, b^{\,\dagger}_{\lambda,j, a}(\p_2) | 0 \ket,  
  \label{EQ:QCDTRIAL2} \\
  | \Psi_4 \ket = & \sum_{i,j,k,l} \, \Lambda_{ijkl} \sum_{\kappa, \lambda, \mu, \nu} \int d\p_{1..4} \, \cG_{\kappa, \lambda, \mu, \nu}(\p_{1..4}) \, a^{\dagger}_{\kappa,i,a}(\p_1) \, b^{\,\dagger}_{\lambda,j,a}(\p_2) \, a^{\dagger}_{\mu,k,b}(\p_3) \, b^{\,\dagger}_{\nu,l,b}(\p_4) | 0 \ket.
  \label{EQ:QCDTRIAL4}
\end{align}
%
%
The term $|\Psi_2 \ket$ is the simplest flavour-$a$ two-body quark-antiquark Fock state, while  $|\Psi_4 \ket$
is a four-body Fock state that contains an additional flavour-$b$ pair,  
$\kappa, \lambda, \mu, \nu$ are the spin indices, and $i, j, k, l$ are the colours indices, while $\Omega$ and $\Lambda$ 
are colour ``wavefunctions" for the two and four components respectively. 
The coefficient functions $\cF$ and $\cG$ must be such that $|\Psi_t \ket$ is an $J^{PC}$ eigenstate.
The constants $C_F$ and $C_G$ are parameters which, modulo normalization, must be determined variationally together with all the parameters contained in $\cF$ and $\cG$. 
\par
The colour index of a quark (anti-quark) transforms in the fundamental (conjugate) representation of $SU(3)$. To find the colour index factors $\Omega$ and $\Lambda$ one must consider group products of $SU(3)$ representations. The dimensions of the group products of $SU(3)$ representation can be determined using the Young tableaux method \cite{bookGeorgi1999}. Appendix B of reference \cite{Thesis} has some details of the derivation of the group products. The two required decompositions are
\begin{gather}
  3 \, \otimes \, \bar{3} = 1 \, \oplus \, 8,
  \label{EQ:ONE_SING} \\
  3 \, \otimes \, \bar{3} \otimes 3 \, \otimes \, \bar{3} = 1 \, \oplus \, 1 \, \oplus \,  8 \, \oplus \, 8 \, \oplus \,  8 \, \oplus \,  8 \, \oplus \, 10 \, \oplus \, 10 \, \oplus \,  27.
  \label{EQ:TWO_SING}
\end{gather}
It is known experimentally that the physically allowed representations are colourless (i.e. zero in each of the eight colour charges; see below) corresponding to the one dimensional representations. The singlet in equation (\ref{EQ:ONE_SING}) provides the colour ``wavefunction" for the single-pair component (\ref{EQ:QCDTRIAL2}). Similarly, the two singlets in (\ref{EQ:TWO_SING}) provide the colour index ``function" for the two-pair component (\ref{EQ:QCDTRIAL4}). This index ``wavefunction" is, again, comprised of delta symbols and, in this case, has to be equally weighted between the two singlets. The properly normalized index functions are 
\begin{gather}
  \Omega_{ij} =  \ds\frac{1}{\sqrt{3}} \, \delta_{ij} \, ,
  ~~~~~~ %
  \Lambda_{ijkl} = \ds\frac{1}{2\sqrt{6}}\left(\de_{ij} \, \de_{kl} + \de_{il} \, \de_{kj}\right) 
  \label{EQ:LAMBDAIJKL}.
\end{gather}
It is easily verified that the trial state (\ref{EQ:QCDTRIAL24}), with the colour index functions given by 
(\ref{EQ:LAMBDAIJKL}), is an eigenstate of the total colour charge operator $Q^a$ ($a=1, ..., 8$) with eigenvalue $0$. The colour charge operator is
\begin{equation}
  Q^a = \int d\x \, \psib_i(x) \, \gamma^0 \, T^a_{ij} \, \psi_j(x) 
  =  \sum_{i,j=1..3} \sum_{s=\pm} \int d\p \; 
 \bigg\{ a^\dagger_{s,i}(\p) \, T^a_{ij} \, a_{s,j}(\p) 
- b^\dagger_{s,i}(\p) \, T^a_{ij} \; b_{s,j}(\p) \bigg\}.
\end{equation}
\par
In order to extract the quark-antiquark potential we must derive the variational equations that follow from the 
variational principle $\delta \bra \Psi_{t} | H - E | \Psi_{t} \ket = 0$. 
The details of the evaluation of the matrix element $\bra \Psi_{t} | H - E | \Psi_{t} \ket$ are given in 
Appendix B of reference \cite{Thesis}.
\par
The variation of the matrix element with respect to both coefficient functions $\cF^\ast$ and $\cG^\ast$ leads to the coupled equations that describe bound states of a meson:
\begin{align}
  & \cF_{\kappa_1, \lambda_1}(\p_{1,2}) \left( \omega_{\p_1}^A + \omega_{\p_2}^A - E \right) = \; \sum_{\kappa_2, \lambda_2} \int d\p_{3,4} \, \left(\cY_{2,2}\right)^{\kappa_2, \lambda_2}_{\kappa_1, \lambda_1}(\p_{1..4}) \; \cF_{\kappa_2, \lambda_2}(\p_{3,4}) \label{EQ:EQRELQCD2} \\
  & \; + \; R \sum_{\kappa_2, \lambda_2, \mu_2, \nu_2} \int d\p_{3..6} \; \bigg\{ \left(\cY_{2,4}\right)^{\kappa_2, \lambda_2, \mu_2, \nu_2}_{\kappa_1, \lambda_1}(\p_{1..6}) + \left(\cC_{2,4}\right)^{\kappa_2, \lambda_2, \mu_2, \nu_2}_{\kappa_1, \lambda_1}(\p_{1..6}) \bigg\} \; \cG_{\kappa_2, \lambda_2, \mu_2, \nu_2}(\p_{3..6}), \nn \\
  & \cG_{\kappa_1, \lambda_1, \mu_1, \nu_1}(\p_{1..4}) \left(\omega^A_{\p_1} + \omega^A_{\p_2} + \omega^B_{\p_3} + \omega^B_{\p_4} - E\right) = \label{EQ:EQRELQCD4} \\ 
  & \; + \; \frac{1}{R} \sum_{\mu_2, \nu_2} \int d\p_{5,6} \; \bigg\{ \left(\cY_{4,2}\right)^{\mu_2, \nu_2}_{\kappa_1, \lambda_1, \mu_1, \nu_1}(\p_{1..6}) + \left(\cC_{4,2}\right)^{\mu_2, \nu_2}_{\kappa_1, \lambda_1, \mu_1, \nu_1}(\p_{1..6}) \bigg\} \; \cF_{\mu_2, \nu_2}(\p_{5,6}) \nn \\
  & \; + \sum_{\kappa_2, \lambda_2, \mu_2, \nu_2} \int d\p_{4..8} \; \bigg\{ \left(\cY_{4,4}\right)^{\kappa_2, \lambda_2, \mu_2, \nu_2}_{\kappa_1, \lambda_1, \mu_1, \nu_1}(\p_{1..8}) + \left(\cQ_{4,4}\right)^{\kappa_2, \lambda_2, \mu_2, \nu_2}_{\kappa_1, \lambda_1, , \mu_1, \nu_1}(\p_{1..8}) \bigg\} \; \cG_{\kappa_2, \lambda_2, \mu_2, \nu_2}(\p_{4..8}), \nn
\end{align}
where $\omega_\p^A = \sqrt{\p^2 + m_A^2}$ and the colour indices have been summed over and the resulting colour factors included in the kernels (relativistic momentum-space
potentials) $\cY_{i,j}$ and $\cC_{i,j}$. The constant $R = \ds\frac{C_G}{C_F}$ is the ratio that specifies the relative contributions of each Fock-space component. The kernel $\cC_{4,4}$ turns out to be zero due to the colour summation as is explained in reference \cite{Thesis}.
\par
Equations (\ref{EQ:EQRELQCD2}) and (\ref{EQ:EQRELQCD4}) are coupled relativistic integral equations for the wavefunctions $\cF$ and $\cG$, which are difficult to solve, even approximately. The function $\cF$ describes a quark-antiquark pair of mass $m_A$, while $\cG$ describes a two quark-antiquark-pair state with quark flavours of mass $m_A$ and $m_B$ respectively. These equations are, in principle, capable of describing the process of string breaking in QCD, {\it i.e.} the transition from a single meson state 
$q_A \, \bar{q}_A$ to a two meson state $q_A \, \bar{q}_A \, q_B \, \bar{q}_B$. The relativistic kinematics are described fully and without approximations. 
However, the dynamics are described approximately by the kernels $\cY_{2,2}$, $\cY_{2,4} = \cY_{4,2}$, 
$\cY_{4,4}$, $\cC_{2,4} = \cC_{4,2}$ and $\cQ_{4,4}$ because of the approximate nature of the trial state (\ref{EQ:QCDTRIAL24}) and the first-order iterative approximation in the reformulation procedure. 
\par
Note that if the two-pair component, $|\Psi_4\ket$ of the trial state (\ref{EQ:QCDTRIAL24}), were not included, {\it i.e.} 
$\cG = 0$ in equation (\ref{EQ:EQRELQCD2}), the only interaction kernel (relativistic momentum-space potential) that would remain is 
${\cY}_{2,2}$. It corresponds to one-gluon exchange, that is the attractive Coulombic potential in the non-relativistic limit.   
\par
Our purpose in this work is to study the quark-antiquark potential inherent in the coupled equations (\ref{EQ:EQRELQCD2}) and  
(\ref{EQ:EQRELQCD4}). For this purpose it is sufficient to consider equation (\ref{EQ:EQRELQCD2}) only. 
Consequently, we do not need to concern ourselves with the kernels $\cY_{4,4}$, and $\cQ_{4,4}$. 
\par
The colour indices have been summed in the kernels $\cY_{2,2}$ and $\cY_{2,4}$ of eq'n (\ref{EQ:EQRELQCD2}) and the results appear as multiplicative factors in front. In the kernel $\cC_{2,4}$ of eq'n (\ref{EQ:EQRELQCD2}), the colour factor is expressed by the contraction $f^{abc}f^{abc}$ of the structure constants. Note that the spin and momentum dependences are still coupled in these relativistic equations whereas the colour dependence is separable and calculated fully. 
\par
The relativistic kernels pertaining to equation (\ref{EQ:EQRELQCD2}) for the function $\cF$ are, explicitly,  
\begin{align}
  & \left(\cY_{2,2}\right)^{\kappa_2, \lambda_2}_{\kappa_1, \lambda_1}(\p_{1..4}) = - \frac{ 2 \, g_s^2 \, m_A^2}{3(2\pi)^3} \, \frac{\de(\p_1 + \p_2 - \p_3 - \p_4)}{\left(\omega^A_{\p_1} \omega^A_{\p_2} \omega^A_{\p_3} \omega^A_{\p_4} \right)^{1/2}} \nn \\
  & \F!\times \Bigg\{\ub(\p_1, \kappa_1) \, \gamma_\mu \, u(\p_3, \kappa_2) \; \vb(\p_4, \lambda_2) \, \gamma^\mu \, v(\p_2, \lambda_1) \left(\frac{1}{(p_4 - p_2)^2} + \frac{1}{(p_3 - p_1)^2}\right)\Bigg\}, 
  \label{EQ:RELY22}\\
  &\left(\cY_{2,4}\right)^{\kappa_2, \lambda_2, \mu_2, \nu_2}_{\kappa_1, \lambda_1}(\p_{1..6}) = \,- \frac{2 \, g_s^2 \, m_A \, m_B}{3(2\pi)^3} \nn \\
  & \F! \times \Bigg\{\frac{\de(\p_1 - \p_3 - \p_5 - \p_6) \, \de(\p_2 - \p_4) \, \de_{\lambda_1 \lambda_2}}{\left(\omega^A_{\p_1} \omega^A_{\p_3} \omega^B_{\p_5} \omega^B_{\p_6} \right)^{1/2}} \, \nn \\
  & \F!\F! \times \ub(\p_1, \kappa_1) \, \gamma_\nu \, u(\p_3, \kappa_2) \; \vb(\p_6, \nu_2) \, \gamma^\nu \, u(\p_5, \mu_2) \left(\frac{1}{(p_1 - p_3)^2} - \frac{1}{(p_5 + p_6)^2}\right) \nn \\
& \F!\H! + \frac{\de(\p_2 - \p_4 - \p_5 - \p_6) \, \de(\p_1 - \p_3) \, \de_{\kappa_1 \kappa_2}}{\left(\omega^A_{\p_2} \omega^A_{\p_4} \omega^B_{\p_5} \omega^B_{\p_6} \right)^{1/2}} \, \nn \\
  & \F!\F! \times \vb(\p_4, \lambda_2) \, \gamma_\nu \, v(\p_2, \lambda_1) \; \vb(\p_6, \nu_2) \, \gamma^\nu \, u(\p_5, \mu_2) \left(\frac{1}{(p_2 - p_4)^2} - \frac{1}{(p_5 + p_6)^2}\right) \Bigg\},
\end{align}
\begin{align}
  & \left(\cC_{2,4}\right)^{\kappa_2, \lambda_2, \mu_2, \nu_2}_{\kappa_1, \lambda_1}(\p_{1..6}) = \nn \\
  &  \F!\, \frac{ f^{abc}f^{abc} \, g_s^4 \, m_A^2 \, m_B}{(2\pi)^6}  \frac{\de(\p_1 + \p_2 - \p_3 - \p_4 - \p_5 - \p_6)}{\left(\omega_{\p_1}^A \omega_{\p_2}^A \omega_{\p_3}^A \omega_{\p_4}^A \omega_{\p_5}^B \omega_{\p_6}^B\right)^{1/2}} \; \frac{1}{(p_1 - p_3)^2} \frac{1}{(p_2 - p_4)^2} \frac{1}{(p_5 + p_6)^2} \nn \\
  &  \F! \times \bigg\{(p_3 - p_1)_\mu \, \ub(\p_1, \kappa_1) \, \gamma_\nu \,  u(\p_3, \kappa_2) \; \vb(\p_4, \lambda_2) \, \gamma^\mu \, v(\p_2, \lambda_1) \; \vb(\p_6, \nu_2) \, \gamma^\nu \, u(\p_5, \mu_2) \nn \\
  & \F!\H! - \, (p_3 - p_1)_\mu \, \ub(\p_1, \kappa_1) \, \gamma_\nu \, u(\p_3, \kappa_2) \; \vb(\p_6, \nu_2) \, \gamma^\mu \, u(\p_5, \mu_2) \; \vb(\p_4, \lambda_2) \, \gamma^\nu \, v(\p_2, \lambda_1) \nn \\
  & \F!\H! + \,  (p_4 - p_2)_\mu \, \vb(\p_4, \lambda_2) \, \gamma_\nu \, v(\p_2, \lambda_1) \; \vb(\p_6, \nu_2) \, \gamma^\mu \, u(\p_5, \mu_2) \;\ub(\p_1, \kappa_1) \, \gamma^\nu \, u(\p_3, \kappa_2) \nn \\
  & \F!\H! - \, (p_4 - p_2)_\mu \, \vb(\p_4, \lambda_2) \, \gamma_\nu \, v(\p_2, \lambda_1) \; \ub(\p_1, \kappa_1) \, \gamma^\mu \, u(\p_3, \kappa_2) \; \vb(\p_6, \nu_2) \, \gamma^\nu \, u(\p_5, \mu_2) \nn \\
  & \F!\H! + \, (p_5 + p_6)_\mu \, \vb(\p_6, \nu_2) \, \gamma_\nu \, u(\p_5, \mu_2) \; \ub(\p_1, \kappa_1) \, \gamma^\mu \,  u(\p_3, \kappa_2) \; \vb(\p_4, \lambda_2) \, \gamma^\nu \, v(\p_2, \lambda_1) \nn \\
  & \F!\H! - \, (p_5 + p_6)_\mu \, \vb(\p_6, \nu_2) \, \gamma_\nu \, u(\p_5, \mu_2) \; \vb(\p_4, \lambda_2) \, \gamma^\mu \, v(\p_2, \lambda_1) \; \ub(\p_1, \kappa_1) \,\gamma^\nu \, u(\p_3, \kappa_2) \bigg\}. 
  \label{EQ:RELC24}
\end{align}
%

\section{Non-relativistic limit and extraction of the inter-quark potential}
In the non-relativistic limit, the spin and momentum dependences of the wavefunctions decouple.
 That is, the functions $\cF$ and $\cG$ can be written as the products
\begin{gather}
  \cF_{\kappa, \lambda}(\p_{1,2}) = \Theta_{\kappa, \lambda} \, F(\p_{1,2}),  ~~~~~~~    %
  \cG_{\kappa, \lambda, \mu, \nu}(\p_{1..4}) = \Xi_{\kappa, \lambda, \mu, \nu} \, G(\p_{1..4})  \label{EQ:SPINFUNC2}.
\end{gather}
The simplest spin configuration to consider is the singlet where the spin eigenvalues are ${\bf S}^2 = S_3 = 0$. 
The corresponding properly normalized spin index functions are
\begin{gather}
  \Theta_{\kappa, \lambda} = \ds\frac{1}{\sqrt{2}}\, \epsilon_{\kappa, \lambda}, ~~~~~~~
  \Xi_{\kappa, \lambda, \mu, \nu} = \ds\frac{1}{3\sqrt{2}} \left(\epsilon_{\kappa \lambda} \, \epsilon_{\mu \nu} + \epsilon_{\kappa \nu} \, \epsilon_{\mu \lambda}\right)
  \label{EQ:SPINXI},
\end{gather}
where $\epsilon$ is the totally-antisymmetric tensor. 
\par
The non-relativistic equation for the function $F$ in the spin singlet configuration can be obtained by multiplying the non-relativistic reduction of equation (\ref{EQ:EQRELQCD2}) by $\Theta$, of equation (\ref{EQ:SPINXI}), and summing over all the spin indices. 
In addition, the functions $F$ and $G$ are expressed in the centre of mass frame, so that the total energy corresponds to the rest mass of the system:
\begin{gather}
  F(\p_{1,2}) = f(\p_1) \; \de(\p_1 + \p_2), \\
  G(\p_{1..4}) = g(\p_{1,2,3}) \; \de(\p_1 + \p_2 + \p_3 + \p_4).  \label{EqG}
\end{gather}
Consequently, one can integrate out one momentum dependence and thus reduce the number of independent momentum variables in the equations. 
This leads to altered interaction kernels which exhibit a degree of skewness (loss of symmetry) in their dependence on the momentum variables. 
\par
The non-relativistic equation for the quark-antiquark system in the singlet configuration of spin and in the centre of mass frame is
\begin{align}
  f(\p_1) \left( \frac{\p^2_1}{m_A} - {\cal E} \right) = \int d\p_3 \, Y_{2,2}(\p_{1,3}) \, f(\p_3) + R \int d\p_{3,4,5} \; C_{2,4}(\p_{1,3,4,5}) \; g(\p_{3,4,5}),
  \label{EQ:EQFORSMALLF}
\end{align}
%
where ${\cal E} = E - 2 \, m_A$ is the non-relativistic energy and all spin indices have been summed. 
 After extensive calculations, the kernels $Y_{2,2}$ and $C_{2,4}$, to order ${\cal O}(m_A^{-2}, m_B^{-2})$, are found to be: 
\begin{align}
  Y_{2,2}(\p_{1,3}) = & \, \frac{4}{3} \frac{g_s^2}{(2 \, \pi)^3} \, \Biggl(\frac{1}{(\p_3 - \p_1)^2}\Biggr) \; \Biggl(1 + \frac{(\p_3 - \p_1)^2}{4 \, m^2_A} + \frac{(\p_1^2 + \p_3^2)}{2 \, m^2_A}\,  \Biggr), \\
  C_{2,4}(\p_{1,3,4,5}) = &
  \, i \, \frac{ f^{abc}f^{abc} \, g_s^4}{4 \, (2\pi)^6 \, m^2_B} \, \frac{1}{(\p_3 - \p_1)^2 \, (\p_4 + \p_1)^2}  \; \Biggl(\frac{13}{2 \, m_A^2} \; \p_1\cdot\p_3\times\p_4 \nn \\
  & \F! - \frac{3 \, (m_B + m_A)}{m_B^2 \, m_A} \; \Bigl(\p_1\cdot \p_5 \times \p_3 \; + \; \p_1\cdot\p_5\times\p_4 \; + \; \p_5\cdot\p_3\times\p_4\Bigr) \Biggr).
\end{align}
The kernel $Y_{2,4}$ has been left out since its leading contribution is of order ${\cal O}(m_A^{-3}, m_B^{-3})$.
\par
The appearance of the factor $i$ in $C_{2,4}$ may seem to be troublesome since the inter-particle potential can not be imaginary. However, the quantity $R$ is not restricted to be real and one can make a choice such that it is purely imaginary to enforce the overall term to be real. 
Even so, there still remains the choice of the phase which leaves the sign in front undetermined. 
\par
We Fourier transform equation (\ref{EQ:EQFORSMALLF}) to coordinates space by multiplying it by $\ds\frac{\e^{-i\, \p_1\cdot\r}}{(2 \, \pi)^{3/2}}$ and integrating over $\p_1$. Such an operation does not lead to an equation where the wavefunction and inter-particle potential stand as separate factors in all terms. In the terms where the decoupling does not occur, one is forced to multiply and divide by $f(\r)$, the Fourier transform of $f(\p_1)$. Subsequently, an ansatz for $f$ must be provided. For the ground state we make the choice
\begin{align}
  f(r) = \sqrt{\frac{1}{\pi \, a^3}} \, \exp{\Bigl(-\frac{r}{a}\Bigr)},{~~~ \rm or,~in~momentum~space,~~~} 
f(p) = \frac{\sqrt{8} \, a^{3/2}}{\pi \, (\p^2 \, a^2 + 1)^2},
  \label{EQ:HYDROGEN}
\end{align}
which is the normalized wavefunction for the ground state of hydrogen (or positronium), with $a$ being the characteristic size.
\par
Similarly, for the function $g(\p_4, \p_5, \p_6)$, in equation (\ref{EqG}), we choose the following variational ansatz 
(cf. reference \cite{E-R JD 2006}):  
\begin{align}
  g(\p_4, \p_5, \p_6) \sim \frac{a^{9/2}}{f^{abc}f^{abc}} \, \frac{(2 \, \pi)^4 \, 2^{3/2} \, \pi}{(\p_4^2 \, a^2 + 1)^2} \, \frac{1}{(\p_5^2 \, a^2 + 1)^2} \, \frac{1}{(\p_6^2 \, a^2 + 1)^4}, 
  \label{EQ:SMALLG}
\end{align}
where the factors in front are included for convenience and the normalization of this wavefunction is absorbed into the definition of $R$. 
Equation (\ref{EQ:SMALLG}) reflects, as in equation (\ref{EQ:HYDROGEN}), a factorized hydrogen-like dependence for each three-momentum variable and is appropriate for the ground state since it contains no angular dependence. 
\par
Upon calculating the Fourier transform of equation (\ref{EQ:EQFORSMALLF}) using the ans\"atze (\ref{EQ:HYDROGEN}) and (\ref{EQ:SMALLG}), the equation in coordinate representation becomes:
\begin{align}
  -\, \frac{\nabla^2}{m^A} \,  f(\x) +  \Big[V_1(x)  + V_2(x)\Bigr] \,  f(\x) = \, {\cal E} \, f(\x), ~~~~~~ x = |\x|.
\end{align}
The $V_1$ contribution to the inter-particle potential is obtained from
\begin{align}
  -  V_1(x) \, f(\x) &= \frac{1}{(2 \, \pi)^{3/2}} \, \int d\p_{1,3} \; \e^{-i\, \p_1\cdot\x} \;  Y_{2,2}(\p_{1,3}) \, f(\p_3) \\ 
  &= \frac{4}{3} \, \frac{\alpha_s}{2 \pi^2}\int d\p_{1,3} \, \e^{- i \, \p_1\cdot\x}\Bigg\{\frac{f(\p_3)}{(\p_3 - \p_1)^2}  
+ \frac{f(\p_3)}{4 \, m^2_A} 
+ \frac{a^3 \, \e^{x/A}}{2 \pi^2 m_A^2}  \, \frac{(\p_1^2 + \p_3^2)}{(\p_3 - \p_1)^2} \, 
\frac{f(\x)}{(\p_3^2 \, a^2 + 1)^2}\,  \Bigg\},     \label{EQ:PSIV1}
\end{align}
\par \ni
where $\alpha_s = \ds\frac{g_s^2}{4 \, \pi}$ is the dimensionless coupling constant of the strong interaction. The last term in the second line of equation (\ref{EQ:PSIV1}) has been multiplied and divided by $f(\x)$, and (\ref{EQ:HYDROGEN}) used to approximate $f$ in coordinate and momentum spaces. The contribution to the potential energy from the last term is non-local and the procedure of multiplying and dividing by $f$ provides a local representation for it. It follows from (\ref{EQ:PSIV1}) that 
\begin{align}
  V_1(x) = -\frac{4}{3} \, \alpha_s \Bigg\{ \frac{1}{x} +  \frac{\pi}{m^2_A} \, \delta(\x) \, + \frac{a^3}{4 \,  \pi^4 \, m^2_A } \, \exp\Bigl(\frac{r}{a}\Bigr) \int d\p_{1,3} \, \frac{\exp(-i \, \p_1 \cdot\r)(\p^2_1 + \p^2_3)}{(\p_3 - \p_1)^2 \, (\p^2_3 \, a^2 + 1)^2} \Bigg\},
\end{align}
where $x = |\x|$ and $\ds\frac{1}{x}$ is the Coulombic ``one-gluon exchange" term, while the remaining terms are the lowest-order relativistic corrections to it. With a little effort, one can reduce the integral in the last line of (\ref{EQ:PSIV1})  to a double quadrature:
\begin{multline}
  V_1(x) = - \frac{4}{3} \, m_A \, \alpha^2_s \; \Bigg\{ \frac{1}{r} + \alpha^2_s \, \pi \, \delta({\bf r}) \\
  + \alpha_s^2 \, \frac{2}{\pi^2 \,A^2 \, r} \, \exp\Bigl(\frac{r}{A}\Bigr) 
\int_0^{\infty} d p_1 \, \sin\Bigg( p_1 \frac{r}{A}\Bigg) \int_0^{\infty} d p_3 \; p_3 \; 
\frac{p_1^2 + p_3^2}{(p^2_3 + 1)^2} \, \ln\bigg|\frac{p_1 + p_3}{p_1 - p_3}\bigg| \Bigg\},
  \label{EQ:V1}
\end{multline}
where $r = |{\bf r}| = |\x| \, m_A \, \alpha_s$ is a dimensionless quark-antiquark separation variable, $A = a \, m_A \, \alpha_s$ is the analogous dimensionless scale parameter and the remaining integrals are expressed in terms of dimensionless momentum variables. 
\par
The $V_2$ contribution to the inter-particle potential, which comes from the $|\Psi_4\ket$ component of the 
trail state (\ref{EQ:QCDTRIAL24}), is given by 
\begin{align}
  - & V_2(x) \; f(\x) =  \frac{R}{(2 \, \pi)^{3/2}} \, \int d\p_{1,3,4,5} \, \e^{-i\, \p_1\cdot\x} \;  C_{2,4}(\p_{1,3,4,5}) \, g(\p_{3,4,5}) \nn \\
  = & \; R \, i \,\frac{\alpha^2_s \, a^6}{m_B^2} \, \e^{x /a}\;  f(\x) \; \int d\p_{1,3,4,5} \, \e^{ - i \, \p_1\cdot\x} 
 \frac{1}{(\p_3 - \p_1)^2} \, \frac{1}{(\p_4 + \p_1)^2} \, \frac{1}{(\p_3^2 \, a^2 + 1)^2} \, \frac{1}{(\p_4^2 \, a^2 + 1)^2} \, \frac{1}{(\p_5^2 \, a^2 + 1)^4}  \nn \\                              
  & \times \, \Biggl(\frac{13}{2 \, m_A^2} \; \p_1\cdot\p_3\times\p_4 - \frac{3 \,  (m_B + m_A)}{m_B^2 \, m_A} \; \Bigl(\p_1\cdot \p_5 \times \p_3 \; + \; \p_1\cdot\p_5\times\p_4 \; + \; \p_5\cdot\p_3\times\p_4\Bigr)\Biggr) 
  \label{EQ:PSIV2}
\end{align}
where, again, it has been multiplied and divided by $f$ and the ans\"atze (\ref{EQ:HYDROGEN}) and (\ref{EQ:SMALLG}) have been used. From this, it follows that the $V_2$ contribution to the potential is
\begin{align}
  V_2(\x) & = \; \mp  \, |R^\prime| \frac{m_A \, \alpha^7_s}{\xi^2 \, A^5} \; \e^{r /A}\; \int d\p_{1,3,4,5} \; \e^{ - i \, \p_1\cdot{\bf r} /A} 
\frac{1}{(\p_3 - \p_1)^2} \, \frac{1}{(\p_4 + \p_1)^2} \, \frac{1}{(\p_3^2 + 1)^2} \, \frac{1}{(\p_4^2 + 1)^2} \, 
\frac{1}{(\p_5^2 + 1)^4} \nn \\         
  & \times \, \Biggl(\frac{13}{2} \; \p_1\cdot\p_3\times\p_4 - \frac{3 \,  (\xi + 1)}{\xi^2} \; \Bigl(\p_1\cdot \p_5 \times \p_3 \; + \; \p_1\cdot\p_5\times\p_4 \; + \; \p_5\cdot\p_3\times\p_4 \Bigr) \Biggr)
   \label{EQ:V2}
\end{align}
where the notation $m_B = \xi \, m_A$ has been introduced and, as before, the necessary substitutions have been made
 to make the integration variables dimensionless. Note that the requirement that $V_2(\x)$ be real means that $\ds R \, i = R \, e^{i \pi / 2} = R'$ must be real, either positive or negative. The proper choice of sign will be discussed below.
%
%
\section{Results for Heavy Quarkonium}
Since we are considering the non-relativistic approximation, the quark flavour  $A$ might be either the charm with the mass 
 $m_A = 1.25\pm0.09 \; \tx{GeV}$ or the bottom quark with the mass $m_A = 4.65\pm0.03 \; \tx{GeV}$ \cite{RPP}. On the other hand, to model a realistic string breaking effect, the lighter quark flavour could be the up or the down quark with approximately identical but ill-defined masses in the neighbourhood of $1.5-7.0 \; \tx{MeV}$ ~\cite{RPP}. Hence, a value of $\xi = 0.001$ is a reasonable estimate of the light-to-heavy quark mass ratio $m_B/m_A$. With $\xi = 0.001$ the first term in equation (\ref{EQ:V2}) can be neglected. 
\par
The value of the coupling constant $\alpha_s$, obtained from expression 
$\ds g_s^2(\lambda) = g_0^2 / [1 + 7g_0^2 \log(\lambda)/(8 \pi^2)]$ (where $g_0$ is the value of $g_s$ at $\lambda = 1$),  
should correspond to the bound state energy of, say, a bottom quark-antiquark system in the singlet spin configuration 
of mass $M_{b{\bar b}} = 9.46 \; \tx{GeV}$ \cite{RPP}. This is the QCD bound state whose potential will be investigated.
 As input one can use the experimentally measured value $\alpha_s(m_Z) = 0.117$ where $m_{Z} = 91.19 \; \tx{GeV}$~\cite{RPP}.  Upon substitution, one obtains $\alpha_s(9.46) = 0.166$, which is the value that we shall use. Note that for the 
bottom quark-antiquark system, one unit of distance in Bohr radius corresponds approximately to $1.54\times10^{-15}\; \tx{m}$. Note also that $V_2 / V_1$ is of order $\alpha_s^5 = 0.00013$.
\par
Equation $(\ref{EQ:V2})$, as it stands, contains two, yet undetermined, variational parameters $A$ and $R$ (or $R^\prime$). The numerical values of these parameters should be obtained from a variational calculation of the matrix element $\bra \Psi_t | \, H - E \, | \Psi_t \ket$. Such a calculation requires great effort which is not undertaken in this work. Instead, the approach shall be to multiply equation $(\ref{EQ:EQFORSMALLF})$ by $f^\ast(\p_1)$ and integrate over the variable $\p_1$:
\begin{align}
  \int d\p_1 \, f^\ast(\p_1) \, f(\p_1) \; \left( \frac{\p^2_1}{m_A} - {\cal E} \right) &= \int d\p_{1,3} \, f^\ast(\p_1) 
\, \cY_{2,2}(\p_{1,3}) \, f(\p_3) \nn \\ &+ R \int d\p_{1,3,4,5} \; f^\ast(\p_1) \; \cC_{2,4}(\p_{1,3,4,5}) \; g(\p_{3,4,5}).  
  \label{EQ:ENERGYEXPRESSION}
\end{align}
Carrying out the indicated integrals (details are given in \cite{Thesis}, Appendix B. section 6.5)
we obtain the following expression for the total non-relativistic energy $\cal E$ of a meson in the singlet spin configuration
 in terms of the variational parameters $A$ and $R$:
\begin{align}
  {\cal E} = E - 2 \, m_A = & \; m_A \, \alpha_s^2 \, \Biggl(\frac{1}{A^2} - \frac{4}{3}\bigg( \frac{1}{A} + \frac{\alpha_s^2(1 + 16 \, \pi^{-2} \, C_1)}{A^3} \bigg)\Biggr) \pm | R^\prime| \, \frac{m_A \, \alpha_s^7}{A^5} \, C_2,
   \label{EQ:ENERGYA}
\end{align}
where, the constant $C_1 \approx1.85044$ emerges from the calculation of the first term on the left hand side of equation $(\ref{EQ:ENERGYEXPRESSION})$. The constant $C_2$:
\begin{align}
  C_2 & = \; \frac{A^5}{m_A \, \alpha_s^7 \,} \, \int d\p_{1,3,4,5} \; f(\p_1) \; C_{2,4}(\p_{1,3,4,5}) \; g(\p_{3,4,5}) \nn \\
  \, & =  \, - \frac{8}{\xi^2} \int d\p_{1,3,4,5} 
\frac{1}{(\p_1^2 +1)^2} \, \frac{1}{(\p_3^2 +1)^2} \, \frac{1}{(\p_4^2 +1)^2} \, \frac{1}{(\p_5^2 +1)^2} \, \frac{1}{(\p_3 -\p_1)^2} \, \frac{1}{(\p_4 + \p_1)^2}  \nn \\     
  \, & ~~~~~ \times  \, \Biggl(\frac{13}{2} \; \p_1\cdot\p_3\times\p_4 - \frac{3 \, (\xi + 1)}{\xi^2} \; \Bigl(\p_1\cdot \p_5 \times \p_3 \; + \; \p_1\cdot\p_5\times\p_4 \; + \; \p_5\cdot\p_3\times\p_4 \Bigr) \Biggr)
  \label{EQ:C2}
\end{align}
is a multidimensional integral expression which one has to solve numerically using, in practice, the Monte Carlo method.
\par
Unfortunately, it turns out that Monte Carlo integration of equation (\ref{EQ:C2}) does not produce reliable results. The troublesome pieces of the integrand are the triple scalar vector products in the numerator. To circumvent this difficulty, one can place an upper bound on the integral by considering upper bounds on the scalar vector products:
\begin{align}
  \p\cdot\q\times{\bf k} = |\p |\, |\q| \, |{\bf k}| \, \sin{\theta} \, \cos{\phi} \leq |\p |\, |\q| \, |{\bf k}|,
  \label{EQ:TSVP}
\end{align}
where $\phi$ is the angle between the vectors $\q$ and ${\bf k}$ and $\theta$ is the angle between the vectors $\p$ and $\q\times{\bf k}$. 
When all scalar vector products are replaced by their upper bound estimates, equation (\ref{EQ:C2}) can be reduced to a triple quadrature in the radial coordinates of the momentum variables:
\begin{align}
   C_2 & \leq \; \frac{24 \, (\xi + 1)}{\xi^4} \, \frac{4 \, \pi^4}{3}\int dp_1 \, dp_3 \, dp_4 \; p_3 \, p_4 \left( p_1 \, p_3 + p_1 \, p_4 + p_3 \, p_4 \right)\nn \\
  \, & \times \, \Biggl(\frac{1}{(p_1^2 + 1)^2} \, \frac{1}{(p_3^2 +1)^2} \, \frac{1}{(p_4^2 +1)^2}  \Biggr) \, 
  \ln\Bigg(\frac{(p_3 + p_1)^2 + \omega^2}{(p_3 - p_1)^2 + \omega^2}\Bigg) \;   \ln\Bigg(\frac{(p_4 + p_1)^2 + \omega^2}{(p_4 - p_1)^2 + \omega^2}\Bigg),
\end{align}
where only the leading terms in the parameter $\xi$ have been kept. To obtain this result, angular integrations resembling those in the calculation of the constant $C_1$ have been employed. The parameter $\omega$ is a regulator which has been inserted to ensure that a numerical evaluation of this triple quadrature converges. 
One must perform many numerical integrations where the value of $\omega$ is reduced in each successive trial. 
If the integral converges then its numerical evaluation should approach a constant value as $\omega$ tends zero. 
In this case, the triple quadrature is found to converge to the value
\begin{align}
  C_2 \leq 1.005\times 10^{16}.
  \label{EQ:NUMERICC2}
\end{align}
\par
\begin{figure}[t]
  \center{
    \includegraphics[scale=0.6]{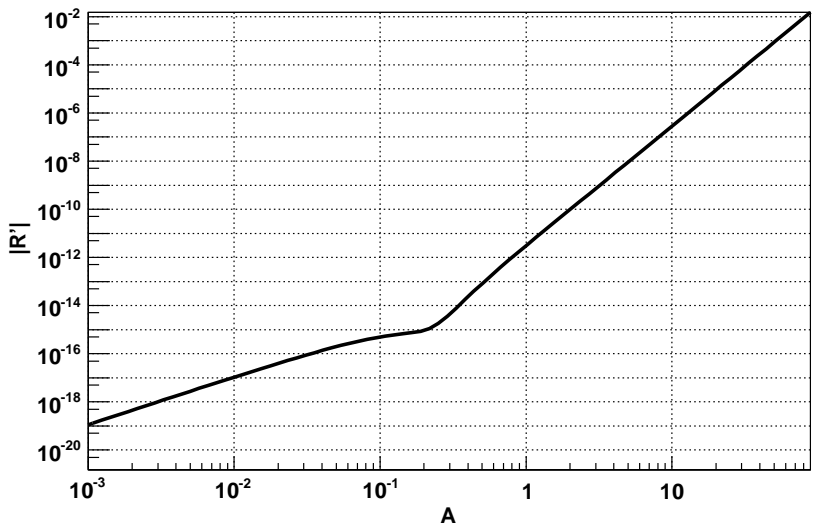}
  }
  \caption{The allowed values for the parameter $A$ and $R^{\prime}$ according to equation (\ref{EQ:ENERGYA}) with the plus sign.}
  \label{FIG:PLOTRA}
\end{figure}

The observed energy ${\cal E}$ of the bottom quark-antiquark state being considered, expressed in units of $m _A\, \alpha_s^2$, is 1.233. Accordingly, one can solve the equation (\ref{EQ:ENERGYA}) to determine the values of $A$ and $R$ which could produce such an energy. Doing so, one finds that only the plus sign in equation (\ref{EQ:ENERGYA}) (correspondingly the negative sign in equation (\ref{EQ:V2})) is realizable for this energy. Figure (\ref{FIG:PLOTRA}) shows a plot of the allowed values of the parameters $A$ and $R^\prime$. 
The optimal value of $A$    
would be $A=1$ if the inter-quark potential were purely Coulombic and the wave-function would be hydrogenic. 
The effect of the confining contribution is to change the shape of the wave function in the large-$r$ domain in which the wave function is rapidly decreasing to zero. Thus, the presence of the confining contribution in the potential 
would not change the value of $A$ substantially, so that the estimate $A=1$ is not unreasonable.

\par
Having estimated the values of the input parameters, we return to the expression for $V_2$, equation (\ref{EQ:V2}). 
This  multidimensional expression can be reduced to the following triple quadrature:
\begin{align}
  V_2(\x) &= \; \mp |R^\prime| \, \frac{m_A \, \alpha^7_s}{A^4} \; \frac{4 \pi^4 \, (\xi + 1)}{\xi^4} \; \frac{\e^{r /A}}{r} 
\int dp_1 \, dp_3 \, dp_4 \; \frac{p_3 \, p_4}{p_1} \, \sin\bigg( p_1 \frac{r}{A} \bigg) \, ( p_1 \, p_3 + p_1 \, p_4 + p_3 \, p_4 )\nn \\
  & \H! \times \, \Biggl(\frac{1}{(p_3^2 + 1)^2} \, \frac{1}{(p_4^2 + 1)^2} \, \ln\Bigg(\frac{(p_3 + p_1)^2 + \omega^2}{(p_3 - p_1)^2 + \omega^2}\Bigg) \, \ln\Bigg(\frac{(p_4 + p_1)^2 + \omega^2}{(p_4 - p_1)^2 + \omega^2}\Bigg) \Biggr) 
  \label{EQ:V2A}
\end{align}
where similar steps have been taken as those in the calculation of the constant $C_2$. Again, the parameter $\omega$ is a regulator which is, ultimately, to be taken to zero. The sign in equation (\ref{EQ:V2A}), which leads to physically meaningful results is the negative one. 
\par
\begin{figure}[t]
  \center{
    \includegraphics[scale=0.64]{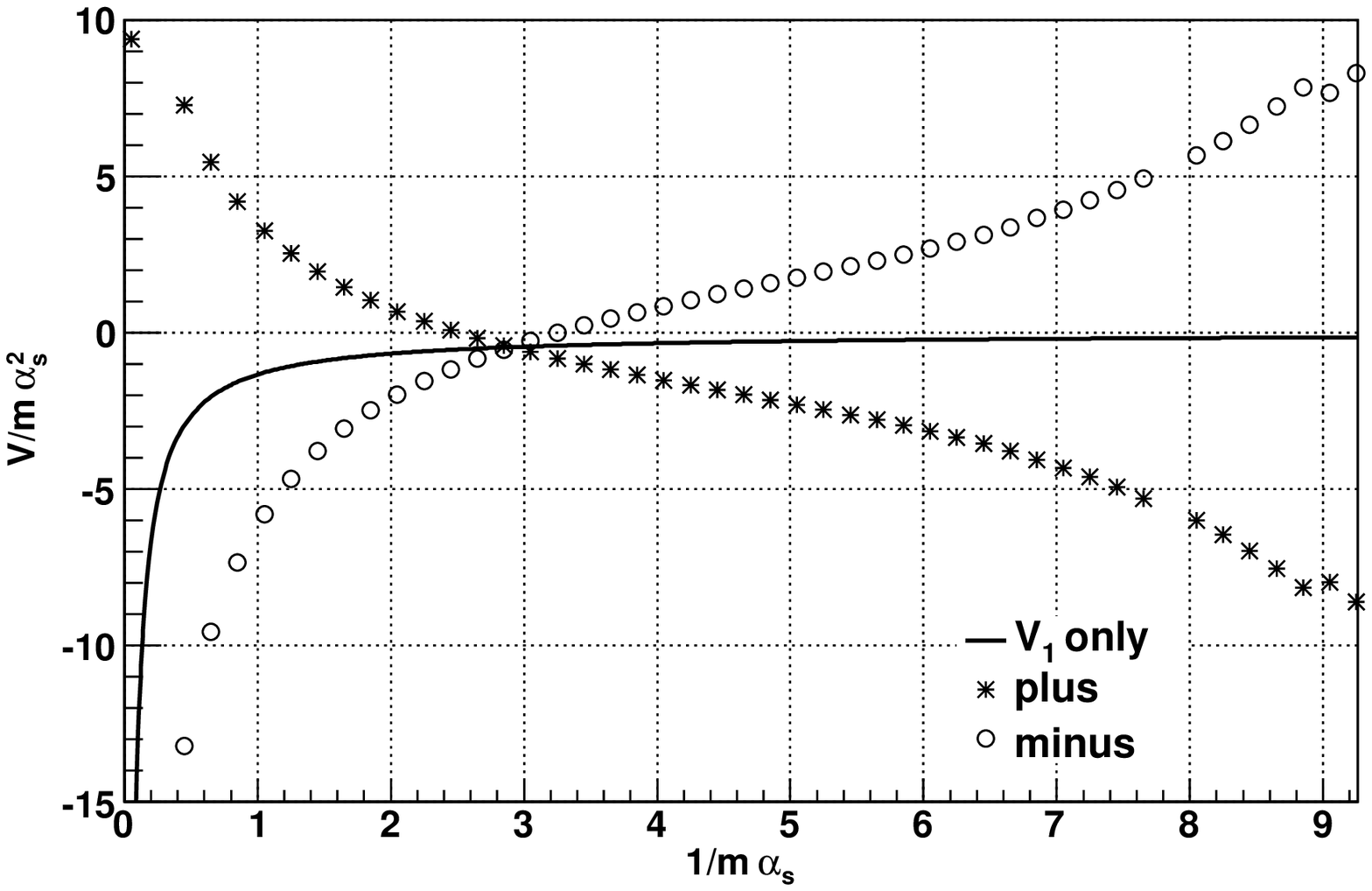}
  }
  \caption{The quark-antiquark potential $V_1 + V_2$ for $A = 1.0$ and $|R| = 3.0\times10^{-12}$. The solid line shows the Coulombic $V_1$ contribution only. The two sets of plotted points corresponds to the two choices of the overall sign in $V_2$ as indicated. The lower curve (labelled ``plus'') is clearly unphysical.}
  \label{FIG:QQPOT1}
  \center{
    \includegraphics[scale=0.64]{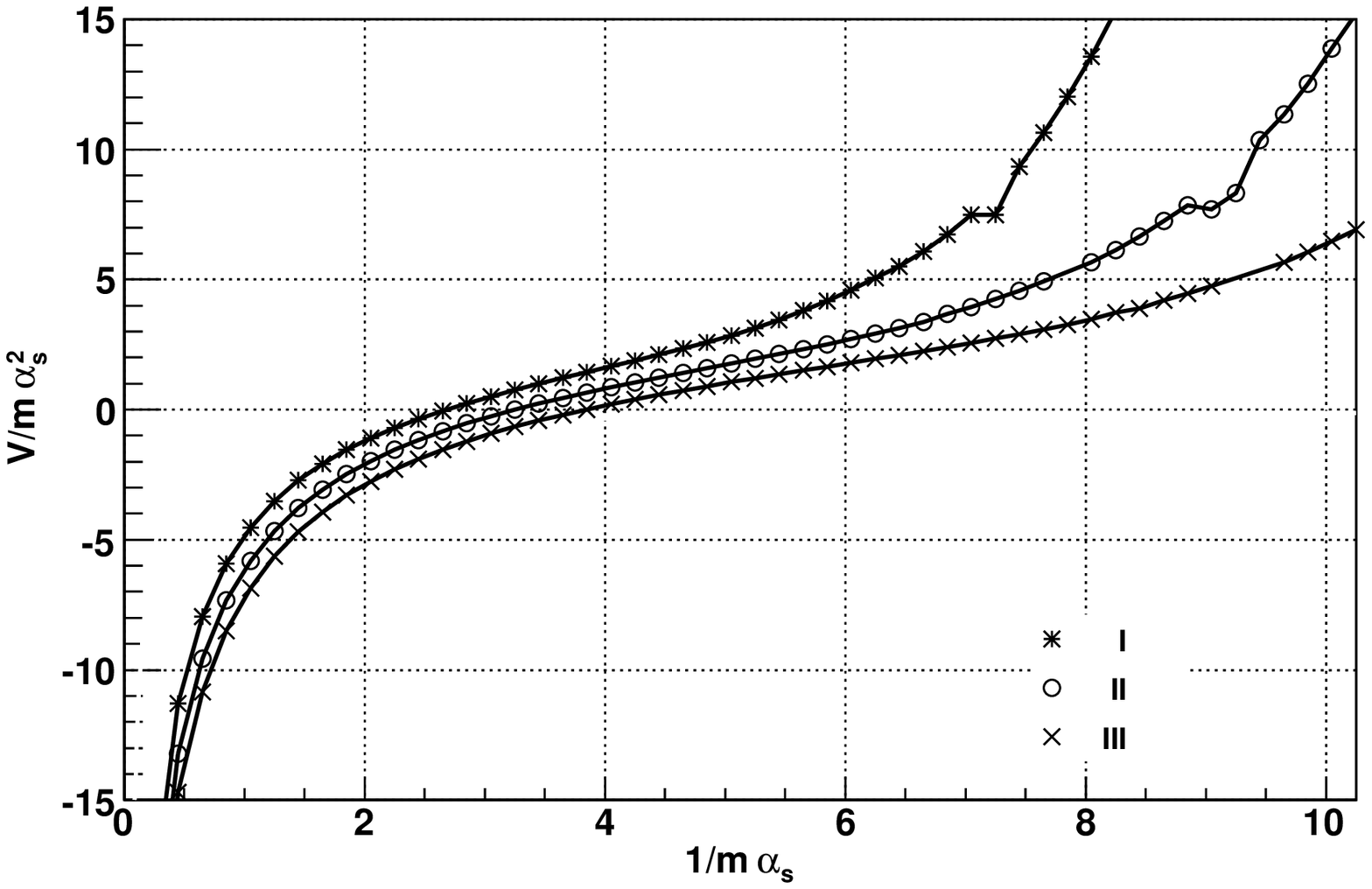}
  }
  \caption{The quark-antiquark potential $V_1 + V_2$ for three choices of the parameters $(A, |R|)$ with the appropriate (-) sign in $V_2$. The choices are I: $(0.8, 9.9\times10^{-13})$, II: $(1.0, 3.1\times10^{-12})$ and III: $(1.2, 7.7\times10^{-12})$. }
  \label{FIG:QQPOT2}
\end{figure}

With the negative sign (-) in equation (\ref{EQ:V2A}), the inter-quark potential $V_1 + V_2$ exhibits a confining character at larger separation distances, whereas the opposite sign (+) yields a potential which does not support stable bound states. 
This is evident from the plots in Figure (\ref{FIG:QQPOT1}), where the inter-particle potential curves corresponding to both signs in equation (\ref{EQ:V2A}) are plotted.  
\par
It is useful to examine the behaviour of the potential for somewhat different values of the parameters $A$ and $R$. To this end, Figure \ref{FIG:QQPOT2} illustrates the quark-antiquark potential for the indicated values of the parameters $A$ and $R$ which are obtained from the curve in Figure \ref{FIG:PLOTRA}. The quark-antiquark potential is substantially altered from its Coulombic behaviour by the $H_{3A}^R$ term of the QCD Hamiltonian (\ref{EQ:QCDHAMREFORM}). The apparent linear segment is characteristic of all three curves (at least in the domain $r \lesssim 7$,  beyond which the numerical results become unreliable). The higher values of $A$ seems to render the linear segment longer. Beyond the separation distances shown on the graph, the points become increasingly scattered. The likely cause of this scattering is due to the difficulty of evaluating accurately the product of the exponential factor $\e^{r/A}$ and the numerical calculation of the triple quadrature in $V_2$, equation (\ref{EQ:V2A}). This scattering can be diminished by requiring greater numerical accuracy in the evaluation of the triple quadrature but, of course, at the expanse of computational time. In any case, the validity of the curves is questionable beyond the linear segment and this is likely due to the limited accuracy in evaluating the triple quadrature. 
\par
It is of interest to extract the string tension $\sigma$ (i.e. the slope of the linear segment) of the quark-antiquark potential and compare it with the value known from lattice gauge calculations. In Greensite's book~\cite{Greensite}, the value obtained from LQCD is quoted to be $\sigma \approx 0.18 \, \tx{GeV}^2$. In Figure \ref{FIG:QQPOT2}, the approximate values of the slopes are $0.16$, $0.13$, $0.11 \; \tx{GeV}^2$ respectively. These derived values are in reasonable agreement with the LQCD calculations. Thus, the present results are gratifying given the approximate estimation of the parameters $A$ and $R$. Also, it must be born in mind that the triple scalar vector products in the integrals (\ref{EQ:V2}) and (\ref{EQ:C2}) are approximated by their upper bounds, cf. equation (\ref{EQ:TSVP}). 
\par
Note that in the approach of this work, the effect of the non-Abelian terms on the inter-particle interactions is to modify the shape of the potential but leave the coupling constant $\alpha_s$ unchanged. 
This is quite different from the (low-order) perturbative S-matrix formalism where the coupling constant $\alpha_s$ becomes energy dependent but the Coulombic shape of the potential remains unchanged. 
Nevertheless, it is interesting to see that the inclusion of a virtual quark-antiquark pair in the trial state (\ref{EQ:QCDTRIAL4}) changes the potential between the heavy valence quark-antiquark pair from purely attractive Coulombic one to one which exhibits linear confinement.
%
%
\renewcommand{\baselinestretch}{1.0}    

\end{document}